\documentclass[11pt,twoside]{article}
\usepackage{asp2010}

\resetcounters

\begin{document}

\title{Overview of the SOFIA Data Processing System:  A generalized system for manual and automatic data processing at the SOFIA Science Center}
\author{R.~Y.~Shuping$^{1,2}$,
R.~Krzaczek$^{2,3}$,
W.~D.~Vacca$^2$, 
M.~Charcos-Llorens$^2$, 
W.~T.~Reach$^2$,  
R.~Alles$^2$, 
M.~Clarke$^2$, 
R.~Melchiorri$^2$, 
J.~Radomski$^2$, 
S.~Shenoy$^2$, 
D.~Sandel$^2$, 
\& E.~B.~Omelian$^4$
\affil{$^1$Space Science Inst., 4750 Walnut St., Suite 250, Boulder, CO  80301, {\tt rshuping@spacescience.org}}
\affil{$^2$USRA-SOFIA, NASA Ames Research Center, N-232, Moffett Field, CA  94035-0001}
\affil{$^3$Carlson Center for Imaging Science, Rochester Inst. of Technology, 54 Lomb Memorial Dr.,
Rochester, NY 15623}
\affil{$^4$Orbital Sciences Corp., TSD, NASA Ames Research Center, Moffett Field, CA 94035-0001}
}

\begin{abstract}
The Stratospheric Observatory for Infrared Astronomy (SOFIA) is an airborne astronomical observatory comprised of a 2.5-meter telescope mounted in the aft section of a Boeing 747SP aircraft.  During routine operations, several instruments will be available to the astronomical community including cameras and spectrographs in the near- to far-IR.  Raw data obtained in-flight require a significant amount of processing to correct for background emission (from both the telescope and atmosphere), remove instrumental artifacts, correct for atmospheric absorption, and apply both wavelength and flux calibration.  In general, this processing is highly specific to the instrument and telescope.  In order to maximize the scientific output of  the observatory, the SOFIA Science Center must provide these post-processed data sets to Guest Investigators in a timely manner.  To meet this requirement, we have designed and built the SOFIA Data Processing System (DPS):  an in-house set of tools and services that can be used in both automatic (``pipeline'') and manual modes to process data from a variety of instruments.  Here we present an overview of the DPS concepts and architecture, as well as operational results from the first two SOFIA observing cycles (2013--2014).  
\end{abstract}

\paragraph{Introduction}

Data collected from the Stratospheric Observatory for Infrared Astronomy (SOFIA)\footnote{\url{http://sofia.usra.edu}} in-flight must undergo significant processing before they can be used in scientific analysis or published.  There are currently four science instruments (SI) available as part of the  Guest Investigator (GI) program\footnote{\url{http://sofia.usra.edu/Science/index.html}}, spanning  optical  to sub-millimeter wavelengths, with additional instruments  to be added in the coming years.  The SOFIA Science Center has defined 4 levels of data products:

\begin{itemize}
\item {\bf Level 1}---Raw data from the instrument but in standardized format (e.g. FITS).
\item {\bf Level 2}---Processed data corrected for instrumental artifacts (e.g. bad pixels, detector non-linearity, flat-fielding).
\item {\bf Level 3}---Flux calibrated data, including telluric correction for spectra.
\item {\bf Level 4}---Higher-order products possibly combining multiple exposures (e.g. mosaics and spectral cubes).
\end{itemize}

GIs are primarily interested in Level~3 and 4 products, but Level~2 data are also useful, especially if the GI wants to apply his or her own flux calibration routines.  {\em In order to maximize the scientific output of  the observatory, the SOFIA Science Center must provide Level~2 and 3 data products to Guest Investigators as soon as possible after the completion of a SOFIA flight series.}  The challenge is to process raw data in a routine fashion with as much automation as possible, while preserving the option for manual reduction in special cases that are not known {\em a~priori}.  In order to achieve this goal, we have designed and built the SOFIA Data Processing System (DPS):  an integrated system that supports both automated (``pipeline'') and manual processing, as well as quality assurance (QA) activities.

\paragraph{System Context}

\begin{figure}[h]
\plotone[scale=0.55]{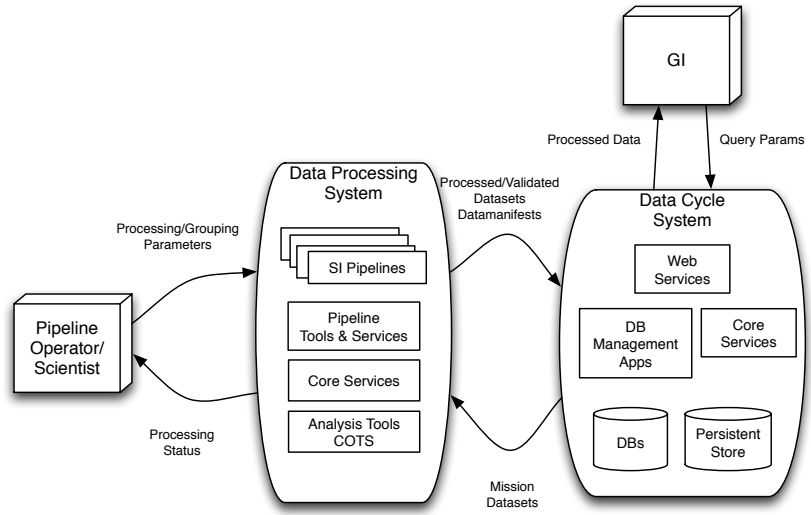}
\caption{\label{fig:DataProcessingContext} System context for the SOFIA Data Processing System (DPS).  An overview of the can be found in \citet{Shuping:2013a}.}
\end{figure}

The SOFIA DPS is a stand-alone system within the SOFIA Science Center network, and is intended to be operated internally by scientists and specialists familiar with data processing for each SOFIA instrument (see Fig.~\ref{fig:DataProcessingContext}).  Raw data is obtained from Data Cycle System (DCS) Persistent Store~\citep{Krzaczek:2014a} post-flight and then processed automatically (or manually if needed) within the DPS.  The processed data products are then transferred back to the DCS for permanent storage and distribution to GIs~\citep{Shuping:2013a}.  The DPS also hosts the needed tools (both COTS and custom built) for carrying out QA assessments and for performing flux calibration.

\paragraph{Tools \& Services}

The DPS is composed of essentially three types of tools and services (shown in Fig.~\ref{fig:DataProcessingContext}):
\begin{itemize}
\item {\bf Instrument Pipelines} process raw (Level 1) instrument data from the aircraft to create Level~2 and 3 data products.  The pipelines are designed to operate only on data-sets from a single observation request (``AOR'').  In principle, these pipelines can be implemented in any reasonable language, but so far all of them have been coded in the Interactive Data Language (IDL).  Instrument pipelines are integrated with our Redux framework~\citep{Clarke:2014} which streamlines automation and provides a common GUI for manual use.  
\item {\bf Pipeline Tools \& Services} provide the necessary infrastructure to automate processing for a whole flight~\citep{Krzaczek:2014} for any pipeline.  The Pipeline Tools \& Services are implemented in C++ and shell scripts.  
\item {\bf Core Services} provide interfaces to the DCS for fetching and storing data from persistent storage and unique tracking identifiers for processed data-sets.  The Core Services are implemented in C++ and shell scripts.
\end{itemize}

\begin{figure}[h]
\plotone[scale=0.5]{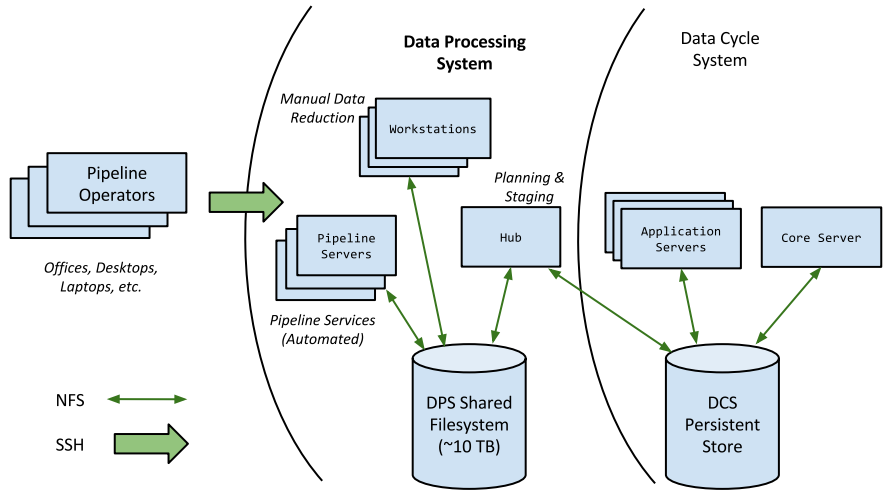}
\caption{\label{fig:DPSNetworkArchitecture} DPS Network Architecture and interface to the DCS.}
\end{figure}

\paragraph{Architecture} DPS hardware consists of a single ``hub'' system coupled to a number of pipeline servers and workstations via NFS (see Fig.~\ref{fig:DPSNetworkArchitecture}).  Automated pipeline services can be deployed to any number of servers (both physical and virtual), allowing the system to scale up as data loads increase (see \citet{Krzaczek:2014} for further details).  All processing software, pipelines, and analysis tools are deployed as shared resources to the hub system so that new servers and/or workstations can be added to the network with minimal configuration.

\paragraph{Standard Operations}

Once a flight is complete and the raw data has been secured in DCS persistent storage, a DPS operator initiates automatic processing for all the raw data on a flight.  The Pipeline Tools first analyze all the data from the flight and group them according to predefined heuristics (usually by AOR).  Then the grouped data-sets are staged to DPS shared storage where they are then processed by the automatic pipelines~\citep{Krzaczek:2014}.  Processing results are stored temporarily on the DPS Shared File-system in well-defined ``preview'' trees with a unique identifier.  This allows all members of the processing team to view and evaluate the results of any processing activity.  Once all data from the flight have been processed, the system sends an email to the operator indicating success or failure for {\em each} group and the location of the preview in shared storage containing the results.  A pipeline scientist or specialist can then log into the system to inspect the data (as part of QA) and re-process  any group manually, as needed.  Once the products have been approved, the entire preview tree is submitted to the DCS for permanent storage, with specific products made available to the appropriate GIs via the DCS Archive Webpage\footnote{\url{https://dcs.sofia.usra.edu/dataRetrieval/SearchScienceArchiveInfoBasic.jsp}}.  Due to the scalable nature of the pipelining system, the number of flights that can be processed simultaneously is limited only by hardware resources.

\paragraph{Operational Results}

The DPS was deployed in July of 2013 and has been used successfully for both SOFIA Observing Cycles 1 and 2, producing over 220~GB of Level 2 and 3 data products  ($>$ 56000 files) .  Currently, the automated pipelines for both FLITECAM and FORCAST can process raw data from a typical flight ($<$ 3~GB) to Level~2 in less than 30~minutes---and often in as a little as 10--15~minutes.  QA inspection and any manual re-processing is usually completed by one person within a few days.  {\em Hence, a full flight of Level~2 data can usually be archived and distributed to GIs within 5 days of flight series completion.} Due to low data loads and the efficiency of the system, we have only needed to deploy a single pipeline server---though we expect to add more servers in the future as SOFIA flight rates increase and especially as instrument detector sizes (and hence data sizes) increase.

\acknowledgements 
\small
The SOFIA Data Processing System is developed and maintained under NASA contract NAS2-97001 to USRA.  

\bibliographystyle{asp2010}
\bibliography{P2-33}

\end{document}